\documentclass[11pt]{article}
\pdfoutput=1
\usepackage{jheppub}
\usepackage{shuffle}
\usepackage{latexsym}
\usepackage{mathrsfs}
\usepackage{amssymb}
\usepackage{amsmath}
\usepackage{cleveref}
\usepackage{slashed,diagbox,stackrel}
\usepackage[colorinlistoftodos]{todonotes}

\usepackage{tikz}
\usepackage{graphicx}
\usepackage{slashed}
\usepackage{adjustbox}
\usetikzlibrary{shapes.geometric, arrows}
\definecolor{light-gray}{gray}{0.95}
\definecolor{light-grayII}{gray}{0.85}
\usepackage{tikz-feynman} 

\newcommand{\C}{\mathbb{C}}

\newcommand{\CP}{\mathbb{CP}}

\newcommand{\R}{\mathbb{R}}
\renewcommand{\P}{\mathbb{P}}

\newcommand{\T}{\mathbb{T}}

\newcommand{\p}{\partial}

\newcommand{\cK}{\mathcal{K}}
\newcommand{\cI}{\mathcal{I}}
\newcommand{\cA}{\mathcal{A}}
\newcommand{\cL}{\mathcal{L}}
\newcommand{\cM}{\mathcal{M}}

\renewcommand{\P}{\mathbb{P}}
\newcommand{\cY}{\mathcal{Y}}

\newcommand{\Pf}{\mathrm{Pf}}

\newcommand{\tr}{\mathrm{tr}}
\newcommand{\sgn}{\mathrm{sgn}}

\newcommand{\rd}{\, \mathrm{d}}

\newcommand{\bt}{\underline{\tau}}

\newcommand{\be}{\begin{equation}\label}
\newcommand{\ee}{\end{equation}}
\newcommand{\bea}{\begin{eqnarray}\label}
\newcommand{\eea}{\end{eqnarray}}

\newtheorem{proposition}{Proposition}[section]
\newtheorem{propn}[proposition]{Proposition}

\newtheorem{lemma}[proposition]{Lemma}

\newcommand{\hook}{{\setlength{\unitlength}{10pt}		
                  ~ \begin{picture}(.833,.8)
                   \put(-.18,.08){\line(1,0){.7}}
                   \put(.5,.08){\line(0,1){.5}}
                   \end{picture}}}
\begin{document}

\title{Lie Polynomials and a Twistorial Correspondence for  Amplitudes}
\author{Hadleigh Frost and Lionel Mason\\
The Mathematical Institute, University of Oxford,\\
AWB, ROQ, Oxford OX2 6GG, United Kingdom}

\abstract{
We review Lie polynomials as a mathematical framework that  underpins  the structure of the so-called double copy relationship between gauge and gravity theories (and a network of other theories besides).  We explain how Lie polynomials naturally arise in the geometry and cohomology of $\cM_{0,n}$, the moduli space of $n$ points on the Riemann sphere up to Mobi\"us transformation.  We introduce a twistorial correspondence  between the cotangent bundle $T^*_D\cM_{0,n}$, the bundle of forms with logarithmic singularities on the divisor $D$ as the twistor space,  and $\cK_n$ the space of momentum invariants of $n$ massless particles subject to momentum conservation as the analogue of space-time. This gives a natural framework for Cachazo He and Yuan (CHY) and ambitwistor-string formulae for scattering amplitudes of gauge and gravity theories as being the corresponding Penrose transform.  In particular we show that it gives a natural correspondence between CHY half-integrands and scattering forms, certain $n-3$-forms on $\cK_n$,  introduced by Arkani-Hamed, Bai, He and Yan (ABHY). We  also give a generalization and more invariant description of the associahedral $n-3$-planes in $\cK_n$ introduced by ABHY.} 


\maketitle

\section{Introduction}
Color-kinematics duality and the double copy \cite{Bern:2008qj, Bern:2010ue} have had a powerful influence on recent developments in scattering amplitudes. They stem from the KLT relations in string theory \cite{Kawai:1985xq} between gravity and Yang-Mills tree-amplitudes and have been devloped as a  tool for the study of multiloop gravity amplitudes and more recently for applications to perturbative classical gravity calculations in connection with gravitational waves \cite{Bern:2019prr}.  Notwithstanding the physical applications, the underlying mathematical framework is perhaps rather surprising even at tree level.  There is little hint of such a double copy structure in standard approaches to perturbation theory of the classical nonlinear theories involved.
The purpose of this article is to develop some of the underpinning mathematical structures.  We  build on observations by Kapranov in an after dinner talk \cite{Kapranov} concerning the relevance of Lie polynomials, both in the double copy and in the Parke-Taylor expressions that pervade the subject.   We also build on the recent work by Arkani-Hamed, Bai, He and Yan \cite{Arkani-Hamed:2017mur} that introduces differential forms in the space of kinematic invariants, $\cK_n$.
We tie them together by means of a  double fibration correspondence that leads to a Penrose-like transform for the   formulae of Cachazo He and Yuan (CHY)  arising from the scattering equations \cite{Cachazo:2013hca,Cachazo:2013iea}. 

The first section provides an elementary review of the theory of Lie polynomials as relevant to this topic, and  expresses standard facts about the double copy in this language.  In particular, the trivalent diagrams of BCJ are a representation of elements $\Gamma\in Lie(n-1)$, the space of Lie polynomials of degree $n-1$, and BCJ numerators $N_\Gamma$ are homomorphisms $N:Lie(n-1)\rightarrow V$ where $V$ is some vector space of polynomials in the polarization data and momenta.

We next review the role played by Lie polynomials in the geometry of the moduli space $\cM_{0,n}$ of $n$ points $\sigma_i$ in $\CP^1$, both in describing the compact cycles in the homology $H_{n-3}(\cM_{0,n}-D)$, which is isomorphic to $Lie(n-1)$, and dually the relative cocycles in $H^{n-3}(\cM_{0,n},D)$, represented by the top degree holomorphic \emph{Parke-Taylor forms}.

With these preliminaries in hand, we study a double fibration between the space of Mandelstam variables, $\cK_n$, and $T^*_D\cM_{0,n}$
\begin{eqnarray}
~& \cY_n=\cK_n\times \cM_{0,n},&(s_{ij},\sigma_j)\nonumber\\
&p\swarrow\qquad\searrow q&\nonumber\\
(s_{ij}),~ \cK_n&&\T= T^*_D \cM_{0,n}, ~ (\tau_i,\sigma_i) . 
\end{eqnarray}
where $p$ forgets the second factor and and $q$ is defined by  the incidence relations
\begin{equation}
\tau_i =E_i(s_{kl}, \sigma_m):=\sum_j \frac{s_{ij}}{\sigma_{ij}}\, , \label{incidence}
\end{equation}
which give the left hand side of the scattering equations. In the language of this correspondence, the CHY formulae are a Penrose transform, being simply the push down of the pullback  of certain forms on $\T$.  We investigate other more geometrical aspects of the Penrose transform.   In particular we show that the top power of the symplectic form $\omega^{n-3}$ provides a correspondence between certain $(n-3)$-forms $w_\Gamma$ on $\cK_n$ that were introduced by ABHY and homology classes in $\cM_{0,n}$.  ABHY use the $w_\Gamma$ as numerators so that given a set of conventional numerators $N$ one can associate a \emph{scattering $(n-3)$-form} $\Omega_N$.  These arise from our double fibration via a Penrose transform also.  Dually, ABHY introduce associahedral $(n-3)$-planes in $\cK_n$ that can be used to convert a scattering form into a conventional amplitude.  We give an improved and extended definition of these, and show how they tie into the geometry of the correspondence.

\section{The double copy and Lie polynomials}

Colour structures for $n$-point amplitudes are degree $n$ invariant polynomials of weight one in each of the $n$ Lie algebra `colours' of the external particles.  These naturally arise in Feynman rules as trivalent Feynman diagrams whose vertices are the structure constants of some unspecified Lie algebra.  If we fix the $n^{\text{th}}$ particle, and an invariant inner product on the Lie algebra, at tree-level, such a polynomial can be realized as the inner product of the $n^{\text{th}}$ colour with the Lie algebra element with a \emph{Lie polynomial} formed by successive commutators of the $n-1$ other colours working through the diagram back from the $n^{\text{th}}$ particle. This section reviews material concerning such colour structures in the language of free Lie algebras and Lie polynomials together with their duality with words formed from permutations of the $n-1$ labels of the first $n-1$ external particles. A classic text on free Lie algebras is \cite{Reutenauer}.

\subsection{A review of words, Lie polynomials and trees}

The space of words, $W({n-1})$, is the $(n-1)!$-dimensional linear span of words $$a=x_{a(1)}x_{a(2)}\ldots x_{a(n-1)},$$ where the letters $x_{a(i)}\in \{x_1,\ldots,x_{n-1}\}$ are all distinct, so that the $a$'s define permutations on $n-1$ letters.  There is a natural bilinear inner product on $W(n-1)$ that is defined on monomials $a$ and $b$ by $$(a,b):=\prod_{i=1}^{n-1}\delta_{a(i)b(i)},$$ i.e., $(a,b)$ is $1$ if $a$ and $b$ are the same word, and $0$ otherwise. A `Lie polynomial' in $W(n-1)$ is any expression formed by taking $n-2$ iterated commutators of the $x_i$. An example is
$$
\Gamma=[x_1,[...,[x_{n-1},x_n]...]] + [x_n,[...,[x_{n-2},x_{n-1}]...]],
$$
where $[x_i,x_j]$ is the commutator, $x_ix_j-x_jx_i$. Let $Lie(n-1)$ be the linear subspace of $W(n-1)$ generated by all Lie polynomials $\Gamma$ of weight $n-1$ in the $n-1$ variables, $x_1, ... , x_{n-1}$, with weight one in each. Every Lie monomial $\Gamma$ defines a rooted trivalent tree decorated by an orientation. We denote this tree also by $\Gamma$. An orientation of the tree can be presented as a planar embedding, where two planar embeddings have the same orientation if they differ from one another by an even number of flips. Thus, for example, the monomial $\Gamma = [2,[[1,3],[5,4]]]$ is associated with the following two planar embeddings, amoung others.

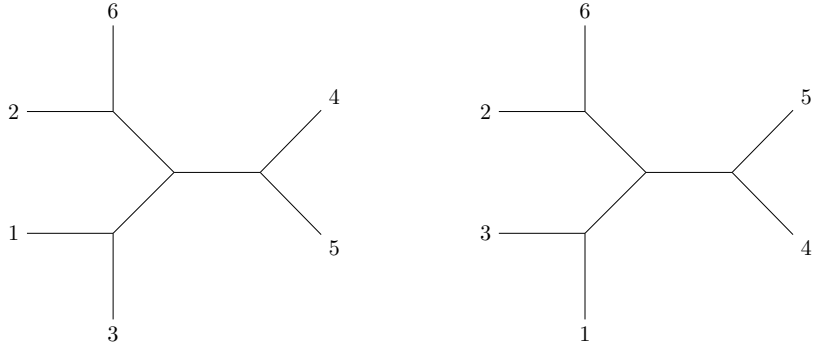
\begin{adjustbox}{width=.7\textwidth
,center}
\begin{tikzpicture}
\begin{feynman}
\vertex (f1);
\vertex [ left=of f1] (b) {\(1\)};
\vertex [below =of f1] (a) {\(3\)};
\vertex [above right = of f1] (i1);
\vertex [above left = of i1] (f2);
\vertex [right = of i1] (e) ;
\vertex [above right=of e] (4) {\(4\)};
\vertex [below right=of e] (5) {\(5\)};
\vertex [above =of f2] (c) {\(6\)};
\vertex [left=of f2] (d) {\(2\)};
\diagram* {
(a) --  (f1) --  (b),
(f1) -- (i1) -- (f2),
(i1) -- (e), (e) -- (4),(e) -- (5),
(c) -- (f2) -- (d),
};
\end{feynman}
\end{tikzpicture}~~~~~~~~~~~~~~~~
\begin{tikzpicture}
\begin{feynman}
\vertex (f1);
\vertex [ below=of f1] (b) {\(1\)};
\vertex [left =of f1] (a) {\(3\)};
\vertex [above right = of f1] (i1);
\vertex [above left = of i1] (f2);
\vertex [right = of i1] (e) ;
\vertex [below right=of e] (4) {\(4\)};
\vertex [above right=of e] (5) {\(5\)};
\vertex [above =of f2] (c) {\(6\)};
\vertex [left=of f2] (d) {\(2\)};
\diagram* {
(a) --  (f1) --  (b),
(f1) -- (i1) -- (f2),
(i1) -- (e), (e) -- (4),(e) -- (5),
(c) -- (f2) -- (d),
};
\end{feynman}
\end{tikzpicture}
\end{adjustbox}

%

The Jacobi identity implies the vanishing of the sum of the three four-point oriented trees corresponding to an $s, t$ and $u$-channel exchange graph.  
  
  \begin{center}
 \begin{tikzpicture}[scale=0.45]
 \draw (0,0) -- (3,0) ;
 \draw (1,0) -- (1,1) ;
 \draw (2,0) -- (2,1) ;
 \node at (-0.4,0) {$1$};
 \node at (3.4,0) {$4$};
 \node at (1,1.4) {$2$};
 \node at (2,1.4) {$3$};
 \node at (4.4,0.5) {{\large $+$}};
 \draw (6,0) -- (9,0) ;
 \draw (7,0) -- (7,1) ;
 \draw (8,0) -- (8,1) ;
 \node at (5.6,0) {$1$};
 \node at (9.4,0) {$4$};
 \node at (7,1.4) {$3$};
 \node at (8,1.4) {$2$};
 \node at (-2,0.5) {{\large $-$}};
 \draw (-7,0) -- (-4,0) ;
 \draw (-5.5,0) -- (-5.5,0.5) ;
 \draw (-5.5,0.5) -- (-5,1) ;
 \draw (-5.5,0.5) -- (-6,1) ;
 \node at (-7.4,0) {$1$};
 \node at (-3.6,0) {$4$};
 \node at (-6,1.4) {$2$};
 \node at (-5,1.4) {$3$};
  \node at (10.4,0.5) {{\large $\quad=0$}};
\end{tikzpicture}
\end{center}

We will denote three Lie polynomials or corresponding graphs that differ only on such a four point subgraph by $\Gamma_s, \Gamma_t$ and $\Gamma_u$, and we will consequently have
\begin{equation}
\Gamma_s+\Gamma_t+\Gamma_u=0.
\end{equation}
The Lie monomial notation is useful for keeping track of the orientations of trees. Recall the inner product $(~,~)$ defined on words. For a Lie monomial $\Gamma$, $(\Gamma,a)$ is the coefficient of $a$ in the expansion of its Lie monomial $\Gamma$. When $\Gamma$ is not planar for the ordering $a$, $(\Gamma,a) = 0$. When $\Gamma$ is planar for the ordering $a$, $(\Gamma, a)$ is the orientation of that planar embedding, either $+1$ or $-1$.  As an application of the notation we can write
\begin{equation}
\Gamma=\sum_a  (\Gamma,a)a \, ,\label{gamma-a}
\end{equation}
which is simply the expansion of the commutators in $\Gamma$.

There are many characterizations of $Lie(n-1)$ as a subspace of $W(n-1)$ \cite{Reutenauer}, and some of these have long been known in the physics literature as relations amoung gauge theory tree amplitudes. For instance, the U(1) decoupling identity is a consequence of Ree's theorem.

\begin{propn}\label{Lie-ree}
(Ree \cite{Ree:1958}) A polynomial $w\in W[n-1]$ is a Lie polynomial iff $(w,a\shuffle b)=0$ for all nontrivial shuffles $a\shuffle b$.\footnote{The shuffle product of $a$ and $b$, $a\shuffle b$, is the sum over ordered permutations of the letters of $a$ and $b$ that preserve the orderings of the letters in $a$ and $b$.}
\end{propn}
This proposition implies that $Lie(n-1)^*$, the dual vector space of $Lie(n-1)$, can be understood as the quotient vector space $W(n-1)/\text{Sh}$, where $\text{Sh}$ is the subspace generated by all nontrivial shuffles.

\begin{lemma}
\label{basis}
(Radford \cite{Radford:1979}) The $(n-2)!$ words of the form $1a$ are a basis for $Lie(n-1)^*$.
\end{lemma}
The direct expansion of $\Gamma_{1a}$ into words is
\begin{equation}
\label{dumb-expansion}
\Gamma_{1a} = \sum_{a\in \bar{u}\shuffle v} (-1)^{|u|}u1v\, ,
\end{equation}
where $|u|$ is the length of $u$.
It follows that
$$
(\Gamma_{1a},1b) = (a,b).
$$
In other words, we have the following.
  
\begin{lemma}
\label{dual-bases}
The Kleiss-Kuijf (KK) basis of $Lie(n-1)^*$ given by words $1a$ is dual to the DDM basis of $Lie(n-1)$ given by combs $\Gamma_{1a}$.
\end{lemma}
An immediate consequence of this Lemma is that any $b+\text{Sh} \in W(n-1)/\text{Sh}$ may be expanded in this basis as
\begin{equation}
\label{dual-lie-expansion}
b+\text{Sh} = \sum_a (\Gamma_{1a},b) 1a + \text{Sh}.
\end{equation}
Dually, given that the combs are a basis for $Lie(n-1)$, a polynomial $w$ is a Lie polynomial iff there is an expansion of $w$ in the combs $\Gamma_{1a}$. Using Lemma \ref{dual-bases}, we find that
\begin{equation}
\label{lie-basis-expansion}
w = \sum (w,1a) \Gamma_{1a}.
\end{equation}
By equation \eqref{dumb-expansion}, we find that
$$
w = \sum (w,1(\bar{u}\shuffle v)) (-1)^{|u|}u1v\, ,
$$
where $\bar{u}$ is the reversal of $u$.
This identity implies the Kleiss-Kuijf relations, which we can restate as a theorem about Lie polynomials.

\begin{propn}\label{Lie-char}
(Kleiss-Kuijf) A polynomial $w\in W[n-1]$ is a Lie polynomial iff the Kleiss-Kuijf relations \cite{KK1989} hold:
\begin{equation}
(w, a1b)=(-1)^{|a|}(w,1\bar a\shuffle \ b)\, ,
\end{equation}
. \end{propn}

\subsection{The geometry of $\cK_n$}

The space of Mandelstam variables is $\cK_n\simeq\R^{n(n-3)/2}$. In coordinates $s_{ij}$, $i,j=1,\ldots , n$ (with $s_{ij}=s_{ji}$, $s_{ii}=0$), $\cK_n$ is the hyperplane given by the equations 
\begin{equation}
\sum_{j=1}^n s_{ij} = 0. \label{conservation}
\end{equation}
For amplitudes, the key geometric structure in $\cK_n$ are the factorization hyperplanes given by $s_I=0$, where $I\subset \{1,2,\ldots,n\}$ and
\begin{equation}
s_I:=\sum_{i,j\in I} s_{ij}=\left(\sum_i k_i\right)^2\, .
\end{equation}
Let $\bar I$ be the complement of $I$, so that $s_{\bar I}=-s_I$ by \eqref{conservation}. Locality states that the only singularities of tree amplitudes are simple poles on these hyperplanes. A further requirement is that a double pole on the intersection of $s_I=0$ and $s_J=0$ occurs in an amplitude only if $I\subset J$ or $\bar J$. It follows that the allowed pole structure of a contribution to an $n$-point amplitude has poles along at most $n-3$ factorization hyperplanes $s_{I_p}=0$ for $p=1,\ldots ,n-3$.  Such choices are in one-to-one correspondence with trivalent (and unoriented) trees.

\subsection{The double copy from biadjoint scalars to gauge and gravity theories}
The double copy principle is that massless $n$-point tree amplitudes for a large web of important  theories, including many gauge and gravity theories,  can be expressed as a double copy in the form
\begin{equation}
\cM=\sum_\Gamma \frac{N_\Gamma \tilde N_\Gamma}{d_\Gamma}\, .
\end{equation}
Here the denominators
\begin{equation}
d_\Gamma=\prod _{r=1}^{n-3} s_{I_r}
\end{equation}
are the propagator factors associated to the graph $\Gamma$ thought of as a Feynman graph. Further, each trivalent diagram $\Gamma$ has a pair of \emph{numerator}  factors   $N_\Gamma$ and $\tilde N_\Gamma$ that are functions of momenta, polarization data, flavour and colour.  Such factors are said to be \emph{local} if they are polynomial, i.e., admit no spurious singularities.  

The key additional feature required to be a BCJ numerator is that $N_\Gamma$ and $\tilde N_\Gamma$ should represent homomorphisms from  Lie polynomials  to some vector space $V$ of functions,
\begin{equation}
N:Lie(n-1)\rightarrow V.
\end{equation}
Thus  for any three graphs $\Gamma_s$, $\Gamma_t$ and $\Gamma_u$ satisfying $\Gamma_s+\Gamma_t+\Gamma_u=0$ as Lie polynomials, we must also have that
\begin{equation}
N_{\Gamma_s}+N_{\Gamma_t}+N_{\Gamma_u}=0\, .\label{num-Lie}
\end{equation}
For this reason, the numerators are not uniquely determined: given a triple $\Gamma_s$, $\Gamma_t$ and $\Gamma_u$ we can perform the shift $\delta(N_{\Gamma_s}, N_{\Gamma_t}, N_{\Gamma_u})=(s,t,u)A$ for any $A\in V$. It also follows that BCJ numerators can be determined from their values on a comb basis $N_{\Gamma_{1a}}$ by
\begin{equation}
N_\Gamma =\sum_{a} (\Gamma ,1a) N_{\Gamma_{1a}}\, .\label{N-basis}
\end{equation}
In the case of Yang mills, the claim of BCJ \cite{Bern:2008qj} is that
\begin{equation}
\cA=\sum_\Gamma\frac{N_\Gamma^{k,\epsilon} c_\Gamma}{d_\Gamma}
\end{equation}
for some \emph{kinematic numerators} $N_\Gamma^{k,\epsilon}$ depending linearly on each polarization vector $\epsilon_i$ and rationally (or even polynomially) on the momenta satisfying \eqref{num-Lie}.  The key nontrivial output of the double copy is that gravity amplitudes are obtained  when $\tilde N_\Gamma=N_\Gamma^{k,\epsilon}$. The same numerators determine both 
the colour-ordered Yang-Mills amplitude with order $a$ is then
\begin{equation}
\cA_a = \sum_\Gamma \frac{N_\Gamma (\Gamma,a)}{d_\Gamma }\label{YM-N}
\end{equation}
and gravity amplitudes by
\begin{equation}
\cM = \sum_\Gamma \frac{N_\Gamma N_\Gamma}{d_\Gamma }\, .
\end{equation}
The most basic theory in this framework is the bi-adjoint scalar theory whose colour ordered amplitudes are given by
\begin{equation}
m(a,b):=\sum_\Gamma \frac{(\Gamma,a)(\Gamma,b)}{d_\Gamma}\, .\label{mab}
\end{equation}
and we can introduce two underlying abstract amplitudes for these theories given by
\begin{equation}
m=\sum_\Gamma \frac{\Gamma\otimes \Gamma}{d_\Gamma}\in Lie(n-1)\otimes Lie(n-1) \,, \qquad m(a)=\sum_\Gamma\frac{(\Gamma,a)\Gamma}{d_\Gamma}\in Lie(n-1) .
\end{equation}
Substituting \eqref{N-basis} into \eqref{YM-N} we obtain Yang-Mills amplitudes in terms of numerators and $m(a,b)$ by
\begin{equation}
\cA_a=(N,m(a))=\sum_b m(a,1b) N_{\Gamma_{1b}}^{k,\epsilon}\, ,\label{A-m-N}
\end{equation}
with a similar form for gravity
\begin{equation}
\cM=(N\otimes \tilde N, m) =\sum_{a,b} m(1a,1b)N_{1a}^{k,\epsilon}\tilde N_{1b}^{k,\tilde\epsilon}\,.
\end{equation} 

We briefly remark that the basic kinematic numerators $N^{k,\epsilon}_\Gamma$ for Yang Mills were obtained in \cite{Fu:2017uzt} and other related ones such as $N^{k,k}_\Gamma$ can be created by setting $k_i\cdot \epsilon_j=0$, and $\epsilon_i\cdot \epsilon_j=k_i\cdot k_j$ or $N^{\epsilon,k,m}_\Gamma$ by taking some components of the polarization vectors to be in a higher dimension to the momenta. We thereby obtain,  amplitudes for a variety of theories as
$$
\mbox{ Amplitudes: }\qquad\cM=\sum_\Gamma \frac{N_\Gamma ^l N_\Gamma^r}{d_\Gamma}
$$
with $N_\Gamma$'s from in table \ref{models}. See \cite{Bern:2019prr} for an up-to-date list of available numerators and their details.

\renewcommand{\arraystretch}{1.8}
\begin{table}[h]
\begin{center}
\begin{tabular}{|c||l|l|l|l|}
  \hline
  \diagbox{$N^l$}{$N^r$}& $N_\Gamma^{k,\epsilon}$ & $N_{\Gamma}^{k,k}$ & $N_{\Gamma}^{k,\epsilon, m}$ & 
   $c_{\Gamma} \mbox{ or } (\Gamma,a) $\\ \hline \hline
  $N_\Gamma^{k,\epsilon}$ & E &  & & 
   \\ \hline 
  $N_{\Gamma}^{k,k}$ & BI & Galileon 
    &  & \\ \hline  
  $N_{\Gamma}^{k,\epsilon,m}$ & $\stackrel[\text{U}(1)^{m}]{}{EM}$ & DBI & $\stackrel[\text{U}(1)^{m}\times \text{U}(1)^{\tilde m}]{}{EMS}$
  & \\
   \hline 
  $c_\Gamma \mbox{ or } (\Gamma,a)$ & YM &  Nonlinear $\sigma$ & $\stackrel[\text{SU}(N)\times \text{U}(1)^{\tilde m}]{}{EYMS}$ & 
  $\stackrel[\text{SU}(N)\times \text{SU}(\tilde N)]{}{\text{Biadjoint Scalar}}$\\ \hline 
 \end{tabular}
 \caption{Theories arising from the different choices of numerators.} \label{models}
\end{center}
\end{table}

We cannot simply invert \eqref{A-m-N} to obtain the $N_{1b}$ as $m(1a,1b)$ is not invertible. Indeed, 
all theories that can be expressed in this double-copy format with one explicit Lie polynomial factor satisfy the fundamental BCJ relations \cite{Bern:2008qj}.  There are many forms of the relations,  one version being, for a word\footnote{This is eq 3.8 of \cite{Cachazo:2012uq} together with the use of the $U(1)$-decoupling identity in the bracketed term.} $a\in W(n-2)$
\begin{equation}
\sum_{bc=a} s_{i,b_{|b|}} m((i\shuffle b)c)  =0\, ,
\mbox{where } \quad s_{b, k}:=\sum_{i=1}^{|b|} s_{b_i\, k}\, .
\end{equation}
Thus  \eqref{A-m-N} determines the $N_{1a}$ only up to the addition of multiples of the BCJ relations.  This freedom can be used to set all but $(n-3)!$ of the $N_{1a}$ to zero, but this is at the expense of requiring numerators that are rational rather than polynomial in the momenta, so that the remaining numerators will then have spurious poles.

\subsection{A note on the Kinematic Algebra}
Given a Lie algebra, $g$, with an inner product, a Lie monomial $\Gamma \in Lie(n-1)$ gives rise to a `colour factor,' $c_\Gamma$, for every n-tuple of Lie algebra elements, $T_1,...,T_n$,
$$
c_\Gamma : = \tr ( \Gamma [T_1,...,T_{n-1}] T_n ).
$$
This means we have a map
$$
c: Lie(n-1) \rightarrow \otimes^n g^*,
$$
and it is a homomorphism, since the $c_{\Gamma}$ clearly satisfy the Jacobi identity. The BCJ kinematic numerators for Yang-Mills, $N^{k,\epsilon}_\Gamma$, likewise satisfy the Jacobi identity leading to a suggestion that they might arise from some \emph{kinematic algebra}, an as yet unideintified  Lie algebra. If that were the case, the double copy would be replacing the Yang-Mills Lie algebra numerator $c_\Gamma$ with kinematic numerators $N^{k,\epsilon}_\Gamma$.  However \eqref{num-Lie}  does not imply that there is a Lie algebra, $g$, such that $N_\Gamma$ is the colour factor for that Lie algebra.
This is clear for example in the case of $(\Gamma,a)$ (and the forms $w_\Gamma$ below). 
Nevertheless, there has been some interesting work to identify such a Lie algebra associated to the kinematic numerators,  \cite{Monteiro:2011pc,Monteiro:2013rya,chen:2019}, \eqref{num-Lie}.  

In general, a homomorphism from $Lie(n-1)$ to a vector space of functions can be given by choosing any $(n-2)!$ such functions, as in equation \eqref{N-basis}. These have been identified in the case of Yang-Mills by various authors by recursion and in particular for Yang Mills in \cite{Fu:2017uzt}.

\section{Trees and words in $\cM_{0,n}$}
In this section we consider the homology and cohomology of $\cM_{0,n}$, the Deligne-Mumford compactification of the space of $n$ distinct points on the Riemann sphere $\CP^1$ up to M\"obius transformations. We first recall the basic properties of $\cM_{0,n}$. $\cM_{0,n}$ has a normal crossing divisor $D$, whose top dimensional strata are the codimension one components $D_I$, for $I\subset \{1,...,n\}$, where the points in $I$ bubble off onto a new $\CP^1$, attached to the first by a node.
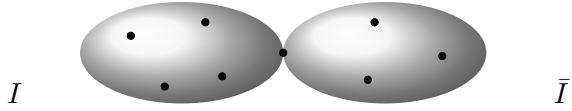
\begin{center}
$I$\qquad\begin{tikzpicture}[scale=4.5]
 \shade[shading=ball, ball color=light-gray] (4.75,1.7) circle [x radius=0.3, y radius=0.15];
 \shade[shading=ball, ball color=light-gray] (5.35,1.7) circle [x radius=0.3, y radius=0.15];
  \draw [fill] (4.6,1.75) circle [radius=.3pt];
 \draw [fill] (4.7,1.6) circle [radius=.3pt];
  \draw [fill] (4.87,1.63) circle [radius=.3pt];
  \draw [fill] (5.3,1.62) circle [radius=.3pt];
  \draw [fill] (5.52,1.69) circle [radius=.3pt];
  \draw [fill] (4.82,1.79) circle [radius=.3pt];
  \draw [fill] (5.32,1.79) circle [radius=.3pt];
  \draw [fill] (5.05,1.7) circle [radius=.3pt];
\end{tikzpicture} \qquad $\bar I$\end{center}
Two such components, $D_I$ and $D_J$, intersect iff $I\subset J$ or $I\subset \bar J$.  Then $D_I\cap D_J$  corresponds to nodal curves with 3 components; for example when $I\subset J$, one containing the points $I$, the second $J-I$ and the third $\bar J$. It follows that the maximal intersections of these $D_I$ are points, $D_\Gamma \in \cM_{0,n}$, given by the intersection of $n-3$ compatible $D_{I_p}$.  Each such point  correspond to a nodal curve with $n-3$ components, each with three points that are either nodes or marked points. Such a tuple of compatible sets $I_p$ defines a trivalent tree, $\Gamma$, with the components corresponding to the vertices and nodes to propagators. The 0-dimensional strata of $D$ are thus in one-to-one correspondence with trivalent (and unoriented) trees.

The complement of the divisor in $\cM_{0,n}$ is an open top cell, $\cM_{0,n}^\#:= \cM_{0,n} - D$, on which we can use simplicial coordinates $\sigma_i$, $i = 1, ... ,n$, with the gauge fixing $(\sigma_1,\sigma_{n-1},\sigma_n) = (0,1,\infty)$. However, in order to study $\cM_{0,n}$ in the neighborhood of the divisor, it is useful to introduce dihedral coordinates, which are a set of cross-ratios of the points. There is one such set of coordinates for every dihedral structure \cite{Brown:2009qja}.  Given an ordering, $a$, the associated dihedral coordinates are the cross-ratios
\begin{equation}
u_{ij}:= (a_ia_{j-1}|a_{i-1}a_j) = \frac{\sigma_{a_i\,a_{j-1}}\sigma_{a_{i-1}\, a_j}}{\sigma_{a_ia_j}\sigma_{a_{i-1}\, a_{j-1}}}\,,
\end{equation}
for $1<i+1<j$ and $\sigma_{ij} = \sigma_i - \sigma_j$. Each such cross-ratio $u_{ij}$ is associated to the cord, $(a_i,a_{j+1})$, of the n-gon labelled by the ordering, $a$. We will also denote $u_{ij}$ by $u_I$, where $I \subset \{1,...,n\}$ is the subset $\{a_i,...,a_{j-1}\}$ or its complement.

If the ordering $a$ is compatible with the set $I\subset \{1,...,n\}$, then the divisor component $D_I$ can be seen in these coordinates as the locus of $u_I = 0$ \cite{Brown:2009qja}.  The points $D_\Gamma$ in the divisor are, in dihedral coordinates, given as follows. Choose any ordering $a$ such that $\Gamma$ is planar for $a$ (i.e. $(\Gamma,a)\neq 0$), with propogators given by the subsets $I_p \subset \{1,...,n-1\}$. In the dihedral coordinates associated to $a$, there are $n-3$ cross-ratios $u_{I_p}$ corresponding to the propogators of $\Gamma$. The point $D_\Gamma$ is then given, in these coordinates, by $u_{I_p}= 0$ for all $p=1,...,n-3$.

The $u_{I_p}$ form a good set of coordinates near $\Gamma$.  Relations between such coordinate systems near different such points are obtained from the non-crossing identity
\begin{equation}
u_{ij}=1-\prod_{(k,l)\in (i,j)^c} u_{kl}\, , \label{non-cross}
\end{equation}
where for $k<l$, $(k,l)\in (i,j)^c$ means that the diagonal $(k,l)$ of the polygon with vertices $\{1,\ldots,n\}$ crosses the diagonal $(i,j)$. 


\subsection{The cohomology of $\cM_{0,n}$ and Parke-Taylors} 
The dimensions of the cohomology groups of $\cM_{0,n}$
are given by the Poincar\'e polynomial\footnote{Arnol'd \cite{Arnold}
computes the Poincar\'e polynomial of the cohomology of the configuration space $M_{n-1}$ of $n-1$ points in $\C$.  This can be obtained inductively via  the fibration $M_{k}\rightarrow M_{k-1}$ with fibre $\C-\{\sigma_1,\ldots,\sigma_{k-1}\}$ giving factors of $(1+(k-1)t)$. To get $\cM_{0,n}^\#$, one needs to take the quotient of $M_{n-1}$ by $\C^*\ltimes \C$, the stabilizer in $PSL_2$ of the point at infinity. Arnol'd's formula for the Poincar\'e polynomial of $M_{n-1}$ must therefore be divided by $1+t$ to obtain the formula here.}
\begin{equation}
P(t) :=\sum_i \dim (H^i(\cM^\#_{0,n}))\,t^i=\prod_{k=2}^{n-2} (1+kt)\, .
\end{equation}
The cohomology ring is generated by the $\rd \log \sigma_{ij}$ in the standard gauge fixing, subject to the quadratic relations 
\begin{equation}
\frac{d\sigma_{ij}}{\sigma_{ij}} \wedge \frac{d \sigma_{jk}}{\sigma_{jk}} +
\frac{d\sigma_{jk}}{\sigma_{jk}} \wedge \frac{d\sigma_{ki}}{\sigma_{ki}} +
\frac{d\sigma_{ki}}{\sigma_{ki}} \wedge \frac{d\sigma_{ij}}{\sigma_{ij}} =0\, . \label{schouten}
\end{equation}
This gives the dimension of $\Gamma(\Omega_D^1)$ as $\sum_{k=2}^{n-2}=n(n-3)/2$ as claimed earlier. It also follows that the top cohomology $H^{n-3}(\cM_{0,n},D)\simeq\Gamma (\cM_{0,n},\Omega^{n-3}_D)$ has dimension $(n-2)!$.  A natural spanning set for 
 $\Gamma(\Omega^{n-3}_D)$ is provided by the Parke-Taylor forms  
\begin{equation}
PT(123\ldots n)=\frac{1}{\mathrm{Vol}SL(2)} \bigwedge_{i=1}^n \frac{d\log \sigma_{i\, i+1}}{2\pi i}\, .
\end{equation}
In our gauge fixing 
\begin{equation}
\frac{1}{\mathrm{Vol}SL(2)}=\frac{(2\pi i)^3\sigma_{1\, n-1}\sigma_{n-1\, n}\sigma_{n1}}{d\sigma_1d\sigma_{n-1} d\sigma_n}
\end{equation}
yielding now for a general choice of permutation $a$ of $1,\ldots ,n-1$
\begin{equation}
PT_a:=PT(a n)=\frac{d^{n-3}\sigma }{\prod_{i=1}^{n-2} 
\sigma_{{a(i)\, a(i+1)}}} \, , \qquad d^{n-3}\sigma:=\frac{1}{(2\pi i)^{n-3}}\bigwedge_{i=2}^{n-2} d\sigma_i\, .
\end{equation}
The $(n-1)!$ Parke-Taylors defined in this way are not linearly independent, because they satisfy the shuffle relations of \ref{Lie-char},
\begin{equation}
\label{eqn:shuffle}
PT_{b\shuffle c}=0,
\end{equation}
for $b,c$ nontrivial, identically\footnote{We thank Carlos Mafra for this reference and also refer the reader to sections 4.1. and 4.2 of  \cite{Mafra:2015vca} for further discussion.} \cite{Cresson:2006}.
Thus, following Proposition \ref{Lie-char}, we deduce that 
\begin{equation}
\label{cohomology-lie-star}
H^{n-3}(\cM_{0,n},D)\simeq\Gamma (\cM_{0,n},\Omega^{n-3}_D) \simeq Lie(n-1)^*.
\end{equation}
Moreover, by Lemma \ref{basis},  one can take a KK basis for $H^{n-3}(\cM_{0,n},D)$, given by the $PT_{1a}$, for all $(n-2)!$ permutations $a$. It follows from equation \eqref{dual-lie-expansion} that we have the following identity.

\begin{lemma}
\label{ddm-type-identity}
The forms $PT_a$ satisfy
$$
PT_a = \sum_b (\Gamma_{1b},a) PT_{1b}.
$$
\end{lemma}

For future reference, observe that we can write $PT_{1a}$ in dihedral coordinates as
\begin{equation}
\label{pt-def}
PT_{1a}=\sgn (a) \,\bigwedge_{i=3}^{n-1} \frac{1}{2\pi i}d\log \frac{ u_{1a_i}}{1- u_{1a_i}}\,,
\end{equation}
since, in the standard gauge fixing,
$$
d\log \frac{ u_{1a_i}}{1- u_{1a_i}} = d\sigma_{a_{i-1}} \frac{\sigma_{1a_i}}{\sigma_{1a_{i-1}}\sigma_{a_{i-1}a_i}}  + d\sigma_{a_i} \frac{1}{\sigma_{a_{i-1}a_i}}.
$$
In fact, for any tree $\Gamma$ compatible with the ordering $a$ we have an associated top form,
\begin{equation}
\label{pt-general}
PT_\Gamma := \bigwedge_{i=1}^{n-3} \frac{1}{2\pi i} d \log \frac{u_{I_p}}{1-u_{I_p}},
\end{equation}
where the $I_p$ are (some ordering of) the subsets defining the propagators of $\Gamma$. In particular, $PT_{\Gamma_{1a}} = PT_{1a}$. $PT_\Gamma$ is not always equal to a standard Parke-Taylor for the given dihedral structure. For the ordering $1a$, an example of a tree, $\Gamma$, such that $PT_{\Gamma} \neq PT_{1a}$ is the `snowflake', $\Gamma = [[[1,2],[3,4]],5]$, for which one finds
$$
PT_\Gamma = PT_{12345} \times (13|24).
$$
However, for a tree $\Gamma$ in which every vertex is attached to at least one external particle, $PT_\Gamma$ does give rise to the ordinary Parke-Taylor for that dihedral structure, and these expressions for $PT_\Gamma$, in dihedral coordinates, were first considered and used by Koba and Nielsen in \cite{Koba:1969rw}.

Note also that $\Gamma \mapsto PT_\Gamma$ is not a homomorphism from $Lie(n-1)$ to $H^{n-3}(\cM_{0,n},D)$, since the $PT_\Gamma$ do not satisfy the Jacobi identity, which in particular implies that equation \eqref{lie-basis-expansion} cannot be used to expand $PT_\Gamma$ in a basis of $PT_a$'s.


\subsection{Homology of $\cM_{0,n}$}
We saw, in equation \eqref{cohomology-lie-star}, that there is an isomorphism of $H^{n-3}(\cM_{0,n},D)$ with $Lie(n-1)^*$. Integration gives a perfect pairing between relative cohomology, $H^{n-3}(\cM_{0,n},D)$, and the homology of the complement of the divisor, $H_{n-3}(\cM_{0,n} - D)$ \cite{Bott:1982}. It follows that the homology $H_{n-3}(\cM_{0,n} - D)$ can be identified (as a vector space) with $Lie(n-1)$.\footnote{There is a free action of the symmetric group $S_n$ on $H_{n-3}(\cM_{0,n}-D)$. As a module for $S_n$, $H_{n-3}(\cM_{0,n}-D)$ is not isomorphic to $Lie(n-1)$, but $Lie(n-1)$ twisted by the sign representation of $S_n$.} In this section, we will review the description of the homology cycles of $\cM_{0,n}-D$.

A point $D_\Gamma$ naturally gives rise to a class in $H_{n-3}(\cM_{0,n}^\#)$ represented by a real half-dimensional torus that surrounds $D_\Gamma$. Explicitly, the cycle can be defined as the locus
\begin{equation}
\label{cycle-def}
C_\Gamma: |u_{I_p}|= \epsilon_p\, ,\qquad p=1,\ldots,n-3
\end{equation}
for some small $\epsilon_p$ and choice of orientation. These cycles were first described in \cite{Cohen:1973}, and they generate the homology, but are not independent.\footnote{The homology of configuration spaces was first computed by \cite{fadell:1962}, but the cycles are  described by Cohen \cite{Cohen:1973}. Cohen was studying the homology of configuration spaces. To apply his results here one uses the isomorphism $\cM_{0,n}-D \simeq \text{Conf}_{n-3}(\mathbb{R}^2-\{0,1\})$.} Integrating a holomorphic top-form with $C_\Gamma$ evaluates the residue at $D_\Gamma$ \cite{Griffiths:1978}.  Lemma 7.1 of \cite{Brown:2009qja} states that $PT_{a}$ can only have a pole at those $D_\Gamma$ which are compatible with the ordering $a$. Moreover, it is clear that we can orient the $C_{\Gamma_a}$ so that
$$
\int_{C_{\Gamma_a}} PT_a = +1,
$$
which leads to the following,
\begin{lemma}
\label{int-pair-2}
The cycles $C_{\Gamma_{1a}}$ represent a basis for $H_{n-3}(\cM_{0,n}^\#)$, dual to the KK basis of $H^{n-3}(\cM_{0,n},D)$:
$$
\int_{C_{\Gamma_{1a}}}PT_{1b} = (1a,1b),
$$
where we have chosen to orient the $C_{\Gamma_{1a}}$ using the $PT_{1a}$ forms so that the residues are $+1$.
\end{lemma}

The relations amoung the top cycles are given by Jacobi-type relations.

\begin{lemma}
\label{lem:jacobi}
(Cohen\footnote{This is a consequence of Theorem 1.2 in \cite{Cohen:1973}, as reviewed in \cite{Cohen:1995}.}) For three trees related by Jacobi, there exists a contraction of the sum of the corresponding cycles, $C_{\Gamma_s}+C_{\Gamma_t}+C_{\Gamma_u}$.
\end{lemma}
An explicit homotopy contracting $C_{\Gamma_s}+C_{\Gamma_t}+C_{\Gamma_u}$ is easy to visualize for $n=4$, since $\cM^\#_{0,4}=\CP^1-\{0,1,\infty\}$, with boundary points  $D_{s}=0$, $D_{t}=1$, and $D_{u}=\infty$. Appropriately oriented, the three small circles around these points add up to zero in homology.

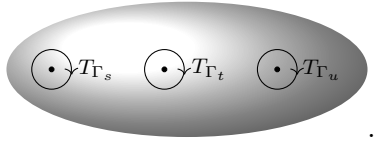
\begin{center}

\begin{tikzpicture} [scale=3]
\shade[shading=ball, ball color=white] (3.6,0)  circle [x radius=0.8, y radius=0.3];
\draw [black, decoration={markings, mark=at position 0.01 with {\arrow{<}}}, postaction={decorate}] (3,0) circle [radius=2.5pt];
  \draw [fill] (3,0) circle [radius=.3pt];
  \node at (3.2,0) {\scriptsize $T_{\Gamma_s}$};
\draw [black, decoration={markings, mark=at position 0.01 with {\arrow{<}}}, postaction={decorate}] (3.5,0) circle [radius=2.5pt];
  \draw [fill] (3.5,0) circle [radius=.3pt];
  \node at (3.7,0) {\scriptsize $T_{\Gamma_t}$};
\draw [black, decoration={markings, mark=at position 0.01 with {\arrow{<}}}, postaction={decorate}] (4,0) circle [radius=2.5pt];
  \draw [fill] (4,0) circle [radius=.3pt];
  \node at (4.2,0) {\scriptsize $T_{\Gamma_u}$}; 
 \end{tikzpicture}.
\end{center}
More generally, for $n>4$, we can exhibit a contraction of $C_{\Gamma_s}+C_{\Gamma_t}+C_{\Gamma_u}$ by making small all $\epsilon$'s in definition of the cycle, equation \eqref{cycle-def}, except for the $\epsilon$ corresponding to the propagator being exchanged. This restricts us to a $\CP^1$ component of $D$, where the same argument made for $n=4$ can be applied.\footnote{The maximum dimension tori appearing in Lemma \ref{lem:jacobi} can be constructed by taking the wedge products of lower dimensional tori. This leads to a natural product in the full homology $H_*(\cM_{0,n}-D)$, under which $H_*$ can be identified with a graded Lie algebra, with graded Jacobi relations holding in all dimensions. \cite{Cohen:1973} The (ungraded) Jacobi relation in Lemma \ref{lem:jacobi} is a special case. \cite{Cohen:1973,Cohen:1995}}

It follows from Lemma \ref{lem:jacobi} and equation \eqref{lie-basis-expansion} that,
$$
C_\Gamma = \sum_{a} (\Gamma,1a)\Gamma_{1a}.
$$
Combining this with Lemma \ref{ddm-type-identity} and Lemma \ref{int-pair-2} implies,
\begin{lemma}\label{int-pairing}
The integration pairing between $C_\Gamma$ and $PT_{a}$ is $$\int_{C_{\Gamma}} PT_{a} = (\Gamma,a).$$
\end{lemma}

\section{A Penrose transform for amplitudes}
Our starting point is the observation that $\cK_n$ can be identified with $H^1(\cM_{0,n},D)$. 
Our `twistor space' for the Penrose transform will be $\T=T^*_D \cM_{0,n}$, the total space of the bundle of holomorphic 1-forms on $\cM_{0,n}$ with logarithmic singularities on $D$. The relationship with $\cK_n$ is given by the isomorphism
\begin{equation}
\cK_n:=H^1(\cM_{0,n},D) \simeq \Gamma({\cM_{0,n}},T^*_D )\, .
\end{equation}
This correspondence can be expressed by considering $d \log $ of the Koba-Nielsen factor \cite{Koba:1969rw}, which is, in the standard gauge fixing,
 \begin{equation}
 KN:=\prod_{i<j} \sigma_{ij}^{s_{ij}}\, , \qquad \sigma_{ij}=\sigma_i-\sigma_j\, .
 \end{equation}
This gives the general section of $\tau\in \Gamma(T^*_D \cM_{0,n}) $ as 
 \begin{equation}
\tau=\sum_iE_id\sigma_i:= \sum_{i<j} s_{ij}d \log \sigma_{ij}\, , \qquad E_i=\sum_j\frac{s_{ij}}{\sigma_{ij}}\, .
\end{equation}
The second equality shows that this is clearly invariant under translations and rescalings of the $\sigma_i$, but  full Mobius invariance (i.e., vanishing when contracted with $\sum_i \sigma_i^2\p_{\sigma_i}$) requires $\sum_i s_{ij}=0$.
Our normalizations $(\sigma_1,\sigma_{n-1},\sigma_n)=(0,1,\infty)$ gives the triviality of $d\log \sigma_{in}$ and $d\log \sigma_{1\,n-1}$ giving the correct dimensionality of the $d\log\sigma_{ij}$ basis of $H^1$. Note  that the equations $E_i=0$ are the \emph{scattering equations}.

To more clearly demonstrate the $d\log$ behaviour on $D$, given a choice of the standard ordering, we can also represent the Koba Nielsen factor as  \cite{Koba:1969kh}
\begin{equation}
KN=\prod_{j>i+1} u_{ij}^{X_{ij}}\,, \qquad X_{ij}=\sum_{i\leq l<m<j} s_{lm}\, . \label{region-KN}
\end{equation}
This gives the useful representation of the general section in terms of  the $n(n-3)/2$ basis $d\log u_{ij}$
\begin{equation}
\sum_iE_id\sigma_i=\sum_{j>i+1 }X_{ij} d \log u_{ij}.\label{tau-u}
\end{equation}
This representation manifests the $d\log$ behaviour on the components of $D$ compatible with this choice of ordering.


\subsection{The double fibration and the CHY formulae}
The twistor correspondence  arises from the following double fibration:
\begin{eqnarray}
~& \cY_n=\cK_n\times \cM_{0,n},&(s_{ij},\sigma_j)\nonumber\\
&p\swarrow\qquad\searrow q&\nonumber\\
(s_{ij}),~ \cK_n&&\T= T^*_D \cM_{0,n}, ~ (\tau_i,\sigma_i) . 
\end{eqnarray}
where $p$ forgets the second factor and and
$q$ is defined by  the incidence relations
\begin{equation}
\tau_i =E_i(s_{kl}, \sigma_m):=\sum_j \frac{s_{ij}}{\sigma_{ij}}\, . \label{incidence}
\end{equation}
A point in $\cK_n$ therefore determines a section $\tau_i=E_i$ of $\T\rightarrow \cM_{o,n}$. 

A special role is played by the zero-section $\T_0$ of $\T$ as it encodes the scattering equations; given generic $s_{ij}$, the section $\tau_i=E_i(\sigma)$ intersects $\T_0$ at the $(n-3)!$ solutions to the scattering equations. We therefore introduce $\bar\delta(\tau)^{n-3}$ to be the $(0,n-3)$-form delta function supported on $\T_0$. In the standard gauge fixing above it can be defined by
\begin{equation}
\bar\delta(\tau)^{n-3}:=\prod_{i=2}^{n-2} \bar\p \frac{1}{2\pi i\tau_i}\, , \qquad \mbox{ where } \qquad\bar\p \frac{1}{2\pi i z}= \delta(\Re z) \delta(\Im z) d\bar z\, .
\end{equation}
More invariantly, this takes values in $(\Omega^{n-3}_D\cM_{0,n})^*$ so that to use it in an integrand, we will need an extra factor with values in  $(\Omega^{n-3}_D\cM_{0,n})^2$.

A first observation is that the CHY formulae can be regarded as examples of a Penrose transform in the sense that the amplitudes are obtained as the pushdown to $\cK_n$ of a pullback of an object from $\T$.  The generic CHY formula takes the form:
\begin{equation}
\cM(s_{ij},\ldots)=\int_{\cM_{0,n}=p^{-1}(s_{ij})} q^*\left(\cI_l \cI_r \,\bar \delta(\tau)^{n-3}\right)
\end{equation}
Here $\cI_l, \cI_r \in \Omega^{n-3}_D\cM_{0,n}$ are CHY half-integrands but also often depending also on polarization data, with the most basic example being $m(a,b)$ when  $(\cI_l,\cI_r)=(PT_a,PT_b)$.
 There is an empirical direct correspondence between choices of $I_{l/r}$ and numerators  $N_\Gamma$ with for example the CHY Pfaffian\footnote{For completeness, the CHY Pfaffian is the reduced Pfaffian of the matrix $M$ 
 $$
M=\begin{pmatrix}A&C\\-C^t&B\end{pmatrix}\, ,\quad 
A_{ij}=\frac{k_i\cdot k_j}{\sigma_i-\sigma_j}, , \quad B_{ij}=\frac{\epsilon_{i}\cdot \epsilon_{j}}{\sigma_i-\sigma_j}\, , \quad C_{ij}=\frac{k_i\cdot \epsilon_{j}}{\sigma_i-\sigma_j}\, , \quad \mbox{ for }i\neq j
$$
 constructed from polarization vectors $\epsilon_{i}$  with $k_i\cdot \epsilon_{i}=0 \ldots$, and $A_{ii}=B_{ii}=0$, $C_{ii}=\epsilon_{i}\cdot \sum k_i/\sigma_{ij}$. This matrix is degenerate, but removing two rows and columns, say $ij$, taking the Pfaffian and dividing by $\sigma_{ij}$ yields a well-defined half-integrand. 
} $\Pf'(M)$ corresponding to the $N^{k,\epsilon}_\Gamma$ described earlier. See \cite{Cachazo:2014xea,Casali:2015vta} for details of half-integrands for other theories and their origins.

\subsection{The geometry of the correspondence}
A generic point $(\tau_i,\sigma_i)$ of $\T$ corresponds to a codimension-$n-3$ plane in $\cK_n$. This plane is the $(n-2)(n-3)/2$ dimensional space of sections that pass through the point. For a point lying in a top-stratum, $D_I$, of the divisor, these planes lie inside the factorisation hyperplane plane $s_I=0$. This follows from the following  combination \cite{Dolan:2013isa} of the scattering equations
\begin{equation}
E_I=\frac{s_I}{2} + \sum_{i\in I, j\in \bar I} s_{ij} \frac{\sigma_{i1}\sigma_{jn}}{\sigma_{ij}\sigma_{1n}}\, .
\end{equation}
and the fact that, assuming $1\in I$ and $n\in \bar I$, the second term vanishes when restricted to $D_I$.

It follows that the point $(\tau_i,\sigma_i)=(0,D_\Gamma)\in \T$ corresponds to the codimension-$n-3$ plane in $\cK_n$ given by the intersection of the planes $s_{I_p}=0$, where $I_p$ are the subsets of $\{1, \ldots ,n-1\}$ corresponding to the momentum flowing through each propagator in $\Gamma$.  We can characterize these planes as being those planes passing through the origin with normal $n-3$-form
 \begin{equation}
 w_\Gamma= \pm\bigwedge_{p=1}^{n-3} d s_{I_p}.
 \end{equation}
In the next section, we will see how the $w_\Gamma$ arise from the double fibration and show how their signs are fixed to make them satisfy the same relations as the $n-3$-forms defined by \cite{Arkani-Hamed:2017mur}.

\subsection{The symplectic form and the holomorphic volume form}
By studying the symplectic volume form $\omega^{n-3}$ on $T^*_D\cM_{0,n}$, we find two elementary consequences of the double fibration. In this section, we describe how the symplectic volume gives rise to a transform between $n-3$ cycles in $\cM_{0,n}-D$ and the $n-3$-forms, equation \eqref{first-forms}, encountered above. In the next section we will describe the associated correspondence between $(n-3)$-planes in $\cK_n$ and $(n-3)$-cycles in $H^{n-3}(\cM_{0,n},D)$. Distinguished classes in $H^{n-3}$ are seen to correspond to the planes defined by \cite{Arkani-Hamed:2017mur}.

The symplectic form on $T^*_D\cM_{0,n}$ can be written explicitly as
$$
\omega = \sum_{i\neq 1,n-1,n} d\tau_i \wedge d\sigma_i,
$$
where $\tau_i$ are the components of $\tau = \sum \tau_id\sigma_i$ in these coordinates. Pulling back $\omega^{n-3}$ to $\cY_n$, we can decompose it into a sum over a basis of $\Gamma(\cM_{0,n},\Omega^{n-3}_D)$ with coefficients given by $n-3$-forms on $\cK_n$. This gives rise to a correspondence between $n-3$-forms on $\cK_n$ and $(n-3)$-cycles in $\cM_{0,n}$ which we explain in this section. In the next section, we describe the associated correspondence between $(n-3)$-planes in $\cK_n$ (`ABHY planes') and $(n-3)$-cocycles in $H^{n-3}(\cM_{0,n},D)$.

Every cycle $C_\Gamma$ in $H_{n-3}(\cM_{0,n}-D)$ defines an $n-3$-form on $\cK_n$,
\begin{equation}
w_\Gamma:=\int_{C_\Gamma} q^* \omega^{n-3} \in \Omega^{n-3}(\cK_n)\, .\label{WG-def}
\end{equation}
It is clear that, for a Jacobi triple of trees,
\begin{equation}
 w_{\Gamma_s}+w_{\Gamma_t}+w_{\Gamma_u}=0. \label{WG-stu}
\end{equation}
In other words, the map $Lie(n-1)\hookrightarrow \wedge^{n-3}\cK_n^*$ given by $\Gamma\mapsto w_\Gamma$ is a homomorphism.

To find an explicit expression for $w_\Gamma$, we use the representation \eqref{tau-u} of $q^*\tau$ in a choice of dihedral coordinates $u_{ij}$ for which $\Gamma$ is planar. This gives
\begin{equation}
q^*\omega=d(q^*\tau)=\sum_{i+1<j} d X_{ij}\wedge\frac{du_{ij}}{u_{ij}}
\end{equation}
 It is then easily seen that integration of $q^*\omega^{n-3}$  over $C_\Gamma$ picks out only those poles that correspond to propagators of $\Gamma$.  The residue thus gives a wedge product $\wedge ds_{I_p}$, with the overall sign determined by the orientation of $C_\Gamma$.
Given this explicit form of $w_\Gamma$, the Jacobi relation, equation \eqref{WG-stu}, can also be understood to follow from the momentum conservation relation, $ds+dt+du=0$ \cite{Arkani-Hamed:2017mur}.  

It follows from equation \eqref{lie-basis-expansion} that $w_\Gamma$ admits the expansion
\begin{equation}
w_\Gamma=\sum_{a\in S_{N-2}} w_{\Gamma_{1a}} (\Gamma,1a).  \label{w-expand}
\end{equation}
Combining this with Lemmas \ref{int-pairing} and \ref{dual-bases}, we find that we can write the pull-back of the symplectic volume form as
\begin{equation}
q^*\omega^{n-3}=\sum_{a\in S_{n-2}} w_{\Gamma_{1a}}  PT_{1a}\, .
\label{w-decomp}
\end{equation}
Although we have used the dual comb and KK bases, this relation follows in any dual basis, because $w_\Gamma$ and $PT_a$ furnish representations of $Lie(n-1)$ and $Lie(n-1)^*$ respectively so that\eqref{w-decomp} is an explicit of writing the Kronecker delta. For example, when $n=4$, $q^*\omega$ can be written in any one of the three KK bases,
\begin{align}
q^* \omega & = -ds_{12} \wedge PT_{213} - ds_{23} \wedge PT_{231} \nonumber\\
& = ds_{12}\wedge PT_{123} + ds_{13}\wedge PT_{132}\nonumber\\
& = -ds_{23}\wedge PT_{321} + ds_{13}\wedge PT_{312}.
\end{align}
More generally, we can rewrite $q^*\omega^{n-3}$ using whatever bases we choose.
\begin{lemma}
\label{w-decomp-general}
The pullback of the symplectic volume form to $\cY_n$ can be written
$$
q^*\omega^{n-3}=\sum_{\Gamma\in H, a\in K} (\Gamma,a)\, w_{\Gamma} \wedge PT_{a}\, .
$$
for a basis $H$ of $Lie(n-1)$, and a basis $K$ of $Lie(n-1)^*$.
\end{lemma}

\begin{figure}[t]
\begin{center}
\begin{tikzpicture}
    \def\r{3}
    \fill[fill=lightgray] (0,0) circle (\r);
    \fill[draw=black,fill=black,thick]
                ({\r*cos(110)},{\r*sin(110)})circle (0.5mm) node[above]{$\mathbf{i}$};
        \fill[draw=black,fill=black,thick]
                ({\r*cos(90)},{\r*sin(90)})circle (0.5mm) node[above]{$\mathbf{i+1}$};
        \fill[draw=black,fill=black,thick]
                ({\r*cos(290)},{\r*sin(290)})circle (0.5mm) node[below]{$\mathbf{j+1}$};
        \fill[draw=black,fill=black,thick]
                ({\r*cos(310)},{\r*sin(310)})circle (0.5mm) node[below]{$\mathbf{j}$};                
        \fill[draw=black,fill=black,thick]
                ({\r*cos(200)},{\r*sin(200)})circle (0.5mm) node[below left]{$\mathbf{n}$};                    
                
\draw[blue,very thick] ({\r*cos(110)},{\r*sin(110)})--({\r*cos(310)},{\r*sin(310)}); 
\draw[red,very thick] ({\r*cos(110)},{\r*sin(110)})--({\r*cos(290)},{\r*sin(290)}); 
\draw[red,very thick] ({\r*cos(90)},{\r*sin(90)})--({\r*cos(310)},{\r*sin(310)}); 
\draw[blue,very thick] ({\r*cos(90)},{\r*sin(90)})--({\r*cos(290)},{\r*sin(290)}); 

\draw[black, dashed, thick] ({\r*cos(200)},{\r*sin(200)})--({\r*cos(130)},{\r*sin(130)}); 
\draw[black, dashed, thick] ({\r*cos(200)},{\r*sin(200)})--({\r*cos(90)},{\r*sin(90)}); 
\draw[black, dashed, thick] ({\r*cos(200)},{\r*sin(200)})--({\r*cos(40)},{\r*sin(40)}); 
\draw[black, dashed, thick] ({\r*cos(200)},{\r*sin(200)})--({\r*cos(310)},{\r*sin(310)}); 
\draw[black, dashed, thick] ({\r*cos(200)},{\r*sin(200)})--({\r*cos(250)},{\r*sin(250)}); 
      
\end{tikzpicture}
\caption{}\label{disk-one}
\end{center}
\end{figure}

\subsection{Associahedral $(n-3)$-planes in $\cK_n$ and forms on $\cM_{0,n}$.}

An alternative way to study the correspondence is to restrict $\omega^{n-3}$ to different $(n-3)$-planes in $\cK_n$. This correspondence is related to the construction in \cite{He:2018pue}, although they discuss different planes and polytopes. Distinguished classes in $H^{n-3}(\cM_{0,n},D)$ correspond to the associahedral planes found in \cite{Arkani-Hamed:2017mur}.

%

An $n-3$-plane in $\cK_n$ is defined up to translation by its tangent form $P \in \wedge^{n-3}\cK_n$. For any such plane, $P$, the symplectic volume form gives rise to an $n-3$-form in $\Gamma(\cM_{0,n},\Omega_D^{n-3})$,
$$
P \hook q^* \omega^{n-3}.
$$
Recall the definition of $PT_\Gamma \in \Gamma(\cM_{0,n},\Omega_D^{n-3})$, given in equation \eqref{pt-general}, which is a top form associated to a dihedral structure, $a$, together with a tree $\Gamma$ that is compatible with that ordering. $PT_\Gamma$ corresponds to a plane $P_\Gamma$ in $\cK_n$ defined as follows. Fix a dihedral structure $a$ and let 
$$D_{I}:=\frac{\p}{\p X_{I}}-\sum _{J\in I^c}\frac{\p}{\p X_{J}},$$
such that we have
\begin{equation}
D_I \log  KN= \log u_I -\sum_{J\in I^c} \log u_J= \log \frac{u_I}{1-u_I}\, ,
\end{equation}
by the non-crossing identity, equation \eqref{non-cross}. It follows that, for a compatible set of propagators $I_p$, defining a tree $\Gamma$ that is compatible with $a$, we can define a plane
$$
P_\Gamma = \bigwedge_{p=1}^{n-3}D_{I_p},
$$
for some ordering of the propagators chosen so that
\begin{equation}
\label{P-first-def}
P_\Gamma \hook q^* \omega^{n-3} = q^* PT_{a,\Gamma}.
\end{equation}
In particular, when we take $P_{a, \Gamma}$ to be the comb, $P_{1a}:=P_{1a,\Gamma_{1a}}$, we recover the Parke-Taylor factors in the standard KK basis,
$$
P_{1a}\hook q^* \omega^{n-3} = q^* PT_{1a}.
$$
For a general $\Gamma$, in dihedral structure $a$, the equations of the plane associated to $P_\Gamma$ can be written as
\begin{equation}
\label{P-def}
X_J + \sum_{r \text{ s.t. } J\in I_r^c} X_{I_r}=\text{const.},
\end{equation}
for all $J$ compatible with the ordering $a$ and not corresponding to a propagator of $\Gamma$. These equations bear no obvious resemblance to those defined in \cite{Arkani-Hamed:2017mur}, but we will see that our planes are the same as theirs. 

First, notice that the $P_a$ satisfy
$$
P_a = \sum (\Gamma_{1b},a)P_{1b},
$$
by Lemma \ref{ddm-type-identity}, and that, moreover,
$$
(w_\Gamma,P_a) = \int_{C_\Gamma} PT_a = (\Gamma, a),
$$
using Lemma \ref{int-pairing}.

In \cite{Arkani-Hamed:2017mur}, the ABHY planes are defined, for the ordering $a=1...n-1$ and the comb $\Gamma_{a}$, to be defined by the $(n-2)(n-3)/2$ equations
$$
s_{ij}=\text{const.},
$$
for $1\leq i < j - 1 < n - 2$. Using the identity, $s_{ij} = X_{ij}+X_{i+1j+1}-X_{ij+1}-X_{i+1j}$ we can verify that, for $I_p$ in $\Gamma_a$,
\begin{equation}
\label{vanished}
D_{I_p} s_{ij} = 0,
\end{equation}
for all $1\leq i < j - 1 < n - 2$, and each $p$. Fixing an $s_{ij}$, one can check this equation for each $p$. The chord $I_p$ is the arc $kn$ for $k=2,...,n-2$. The five cases to check are (i) $k\leq i$, (ii) $k=i+1$, (iii) $i+1<k<j$, (iv) $k=j$, and (v) $k>j$. That equation \eqref{vanished} vanishes in these five cases is easily seen from Figure \ref{disk-one}. Conversely, we can define the ABHY plane for the ordering $a$ and the comb $\Gamma_a$ to be given by the tangent form
\begin{equation}
P_{a }=\bigwedge_{i=2}^{n-2}\left( \frac{\p}{\p s_{a_{i-1}\, a_i}}-\frac{\p}{\p s_{a_i\,a_{i+1}}}\right)\, .
\end{equation}
For each factor in $P_a$ we can check that it annihilates the expressions in equation \eqref{P-def}. We have the following property
\begin{equation}
\left( \frac{\p}{\p s_{i-1\, i}}-\frac{\p}{\p s_{i\,i+1}}\right)X_{jk}=\begin{cases} 1, \quad i=k-1,\\
-1\, , \quad i=j,\\
0\, , \quad \mbox{otherwise.}
\end{cases}
\end{equation}
For fixed $i$, there are five cases to check, and these are shown in Figure \ref{disk-two}. By these arguments, we conclude that

\begin{lemma}
\label{abhy-planes-lemma}
The planes $P_{a,\Gamma}$ defined by the correspondence, equation \eqref{P-first-def}, are the ABHY planes when $\Gamma_a$ is the comb for the ordering $a$.
\end{lemma} 

\begin{figure}[t]

\begin{center}
\begin{tikzpicture}
    \def\r{3}
    \fill[fill=lightgray] (0,0) circle (\r);
    \fill[draw=black,fill=black,thick]
                ({\r*cos(110)},{\r*sin(110)})circle (0.5mm) node[above]{$\mathbf{i}$};
        \fill[draw=black,fill=black,thick]
                ({\r*cos(90)},{\r*sin(90)})circle (0.5mm) node[above]{$\mathbf{i+1}$};               
        \fill[draw=black,fill=black,thick]
                ({\r*cos(240)},{\r*sin(240)})circle (0.5mm) node[below]{$\mathbf{n}$};                    

\draw[blue, very thick] ({\r*cos(240)},{\r*sin(240)})--({\r*cos(110)},{\r*sin(110)}); 
\draw[red, very thick] ({\r*cos(240)},{\r*sin(240)})--({\r*cos(90)},{\r*sin(90)}); 

\draw[black, dashed, thick] ({\r*cos(160)},{\r*sin(160)})--({\r*cos(110)},{\r*sin(110)}); 
\draw[black, dashed, thick] ({\r*cos(170)},{\r*sin(170)})--({\r*cos(90)},{\r*sin(90)}); 
\draw[black, dashed, thick] ({\r*cos(180)},{\r*sin(180)})--({\r*cos(320)},{\r*sin(320)}); 
\draw[black, dashed, thick] ({\r*cos(110)},{\r*sin(110)})--({\r*cos(340)},{\r*sin(340)}); 
\draw[black, dashed, thick] ({\r*cos(90)},{\r*sin(90)})--({\r*cos(350)},{\r*sin(350)}); 
      
\end{tikzpicture}
\caption{}\label{disk-two}
\end{center}

\end{figure}

Moreover, the arguments that led to this Lemma can be applied to the case of general $\Gamma$, which leads to the following

\begin{lemma}
\label{abhy-planes-lemma-2}
The planes $P_{a,\Gamma}$ defined by the correspondence, equation \eqref{P-first-def}, are given by the $(n-2)(n-3)/2$ equations
$$
s_{ij} = \text{const.},
$$
for all chords $(ij)$ compatible with the ordering that are not one of the $n-3$ propagators of $\Gamma$.
\end{lemma} 

It is a consequence of the results in \cite{Arkani-Hamed:2017mur} that the planes in Lemma \ref{abhy-planes-lemma-2} are ABHY planes, in the sense that they cut out associahedra in the same way as the original planes in \cite{thomas:2018}.

\subsection{Scattering forms and CHY}
The scattering forms of \cite{Arkani-Hamed:2017mur} are defined by the following sum over trees,
\begin{equation}
\Omega_a=\sum_{\Gamma}\frac{w_\Gamma  (a,\Gamma)}{d_\Gamma} \, ,\label{old-scat-form}
\end{equation}
where $d_\Gamma=\prod_{p=1}^{n-3} s_{I_p}$ and  $I_p$ are the propagators of $\Gamma$. The $(a,\Gamma)$-factor reduces the sum to one over trees that are planar for the ordering $a$.

The symplectic form discussed above gives rise to the ABHY scattering forms on $\cK_n$ from the Dolbeault formula 
\begin{equation}
\Omega_a = \int_{p^{-1}(s_{ij})} q^* \left[ \bar\delta^{n-3}(\bt) \wedge \omega^{n-3} PT_a \right],\label{new-scat-form}
\end{equation}
where 
\begin{equation}
\bar\delta^{n-3}(\bt)=\bigwedge_{i=1}^{n-3} \bar\delta(\tau_i)\, , \qquad \bar \delta(z)=\frac{1}{2\pi i} \bar\p\frac{1}{z}\, .
\end{equation}
To see that equation \eqref{new-scat-form} is equal to equation \eqref{old-scat-form}, we use Lemma \ref{w-decomp-general} to rewrite equation \eqref{new-scat-form} as
$$
\Omega_a = \sum_b w_{\Gamma_{1b}} \int_{p^{-1}(s_{ij})} q^* \left[ \bar\delta^{n-3}(\bt) \wedge PT_{1b} PT_a \right].
$$ 
We recognize that the integral is the CHY formula \cite{Cachazo:2013hca, Cachazo:2013iea}
 for $m(a,b)$, defined in equation \eqref{mab}. It follows that equation \eqref{new-scat-form} can be expanded as
$$
\Omega_a = \sum_b \frac{w_{\Gamma_{1b}} (\Gamma,1b)(\Gamma,a)}{d_\Gamma},
$$ 
which is equal to \eqref{old-scat-form} by equation \eqref{w-expand}. We conclude that

\begin{proposition}
The ABHY scattering form $\Omega_a$ is given by equation \eqref{new-scat-form}.
\end{proposition}

More generally, given any CHY half-integrand $\cI$, we can define an ABHY scattering form
\begin{equation}
\Omega_{\cI} := \int_{p^{-1}(s_{ij})} q^* \left[ \bar\delta^{n-3}(\bt) \wedge \omega^{n-3} \,\cI\right].
\end{equation}
This form admits an expansion
$$
\Omega_{\cI} = \sum_{a} w_{\Gamma_{1a}} \int_{p^{-1}(s_{ij})} q^* \left[ \bar\delta^{n-3}(\bt) PT_{1a} \cI\right].
$$
The CHY integral as given is not changed by adding to $\cI$ a term that vanishes on the support of the scattering equations. One can regard $\cI$ as a representative of a twisted cohomology class, as explained in \cite{Mizera:2017rqa}. 


All scattering forms, $\Omega_{\cI}$, obtained in this way are projective. Let $\Upsilon$ be the Euler vector field on $\cK_n$,
$$
\Upsilon = \sum_{i<j}s_{ij}\frac{\p~}{\p s_{ij}}.
$$
We can lift $\Upsilon$ trivially to $\cY_n$ using the product structure.  On objects pulled back from $\T$ it then acts by
\begin{equation}
\Upsilon 
= \sum_i \Upsilon(\tau_i)\frac{\p~}{\p\tau_i}=\sum_i\tau_i\frac{\p~}{\p\tau_i}.
\end{equation}
Contracting $\Upsilon$ into $\Omega(a)$ we find
$$
\Upsilon \hook \Omega_{\alpha}^{(n-3)} = \int q^* \left[ \left(\sum_i \tau_i \frac{\p~}{\p\tau_i}\right)\hook \bar\delta^{n-3}(\bt) \wedge \omega^{n-3} PT_a \right] = 0,
$$
on account of the delta functions in the integrand.

\begin{lemma}
The scattering form $\Omega_{\cI}$ is a projective form on $\P\cK_n$.
\end{lemma}
 
Finally, amplitudes are obtained by restricting the scattering forms $\Omega_{\cI}$ to ABHY planes, as in \cite{Arkani-Hamed:2017mur}. In fact, the restriction of $\Omega_{\cI}$ to $P_a$ evaluates to give the CHY formula for the integrands $PT_a$ and $\cI$, as follows from the definition of $P_a$ (equation \eqref{P-first-def} and subsequent lines). Moreover, for a general cohomology class
$$
N := \sum_a N_{1a}PT_{1a} \in H^{n-3},
$$
we have an associated plane, $P_N := \sum_a N_{1a} P_{1a}$. Pairing this with the general scattering form $\Omega_{\cI}$ gives the CHY amplitude for integrands $N$ and $\cI$.

\section{Discussion}
We have seen that Lie polynomials underpin the color-kinematics and double copy framework of BCJ.  We have reviewed the classical fact that the top homology of $\cM_{0,n}-D$ is isomorphic to the Lie polynomials, $Lie(n-1)$, and shown that there is a natural correspondence between $T^*_D\cM_{0,n}$ and $\cK_n$ under which the CHY integral formulae can be understood as a Penrose transform.  This can be extended to a transform between the holomorphic Liouville form and the differential forms $w_\Gamma$ introduced by \cite{Arkani-Hamed:2017mur}, and also between CHY half-integrands and the scattering forms introduced by \cite{Arkani-Hamed:2017mur}.

One underlying question in the subject is whether there is a kinematic algebra underpinning the kinematic numerators $N_\Gamma^{k,\epsilon}$.  Although we have seen that the color factors of Lie algebras can provide such numerators, we have also seen many examples of numerators satisfying the Jacobi identity that do not arise as color factors for a Lie algebra: for example, $(\Gamma,a)$ and $w_\Gamma$ and so on.  In other words, the existence of a homomorphism from $Lie(n-1)$ to some vector space does not of itself determine a Lie algebra.

The basic results in this paper can be taken further to yield natural recursions in field theory, which lead to both Lie polynomial and ABHY-form based proofs of the known properties of the field theory momentum kernel and of kinematic numerators. The momentum kernel can be studied also in $T^*\cM_{0,n}$, where it arises in the CHY treatment of KLT orthogonality \cite{Cachazo:2013gna}.  It also seems likely that the framework will naturally extend to loop integrands in the context of nodal spheres following the logic of \cite{Geyer:2015bja,Geyer:2015jch,Geyer:2016wjx,Geyer:2017ela,he:2018}.

The correspondence we have described is suggestive of the naive explicit  formula
\begin{equation}
N^{\cI_l}_\Gamma=\int_{C_\Gamma}\cI_l\, ,
\end{equation}
for numerators in terms of CHY half-integrands.  However, such formulae fail for the CHY Pfaffian that one would expect to give the basic Yang-Mills kinematic numerators; the formula is compromised by the interdependence between in particular reduced Pfaffians and scattering equations. This equation is shown to be invalid as written  but a slightly different formulation in a similar spirit is shown to work  when $\cI_l$ represents a cohomology class in the context of twisted cohomology by Mizera \cite{Mizera:2019blq}. 

We remark that the twisted cycle formulation of string integrals in \cite{Mizera:2019gea} naturally arises in the context of the holomorphic geometric quantization of $T^*_D\cM_{0,n}$.  To carry out geometric quantization one introduces the line bundle $\cL\rightarrow T^*_D\cM_{0,n}$ with connection $\nabla=d+ \alpha'\tau $  where $\tau=\sum_i \tau_i d\sigma_i$ is the canonical 1-form (symplectic potential) and $\alpha'$ plays the role of Planck's constant. Polarized wave functions should be independent of $\tau_i$.  On pull back to the correspondence space, $\cY_n$, the connection $\nabla $ becomes the standard twisted exterior derivative associated to the Koba-Nielsen factor. 
Such a quantization of $T^*\cM_{0,n}$ perhaps most naturally arises from ambitwistor-string path-integral \cite{Mason:2013sva}, where the Pfaffian half-integrand for kinematic numerators arises from the RNS spin field path-integral.
	
There are many further connections to be followed up; we briefly mention the delta algebras of \cite{Cachazo:2018wvl} and the formulations of colour-kinematics duality in \cite{Chen:2019ywi, Reiterer:2019dys}.

\section{Acknowledgements}
It is a pleasure to acknowledge informative conversations and email exchanges with Nima Arkani-Hamed, Nick early, Yvonne Geyer, Carlos Mafra, Sebastian Mizera, Ricardo Monteiro, Michael Reiterer and Oliver Schlotterer and the hospitality of the CMSA at Harvard. HF would like to acknowledge support from ERC grant Galois theory of periods and applications 724638.  While we were writing up this work we became aware of the parallel work of Sebastian Mizera \cite{Mizera:2019blq} that has some overlap with this---we are grateful to him for letting us see a preview of his work before we finished this paper.

\bibliography{twistor-bib}  
\bibliographystyle{JHEP}

\end{document}